\documentclass{article}
\usepackage{amsmath,amsfonts,epsfig,amssymb}
\usepackage{natbib}

\newcommand{\D}{{\rm d}}
\newcommand{\mean}[1]{ \left\langle #1 \right\rangle}
\newcommand{\NmV}{\ensuremath{{\cal N}^{(m)}(V)}}
\newcommand{\NmtV}{\ensuremath{{\cal N}^{(m)}_{\Delta\vartheta}(\vartheta,V)}}

\newcommand{\E}[1]{\mathbb{E}\left[#1\right]}

\newcommand{\Var}[1]{\mathbb{V}{\rm ar}\left[#1\right]}
\newcommand{\VarS}[1]{\mathbb{V}{\rm ar}\left\{#1\right\}}

\newcommand{\MeanS}[1]{\mathbb{M}{\rm ean}\left\{#1\right\}}
\newcommand{\KurtS}[1]{\mathbb{K}{\rm urt}\left\{#1\right\}}
\newcommand{\KurtO}[1]{\mathbb{K}{\rm urt}\,#1}
\newcommand{\ti}{\ensuremath{{\rm TI}}}
\providecommand\upi{\pi}

\begin{document}
 
\title{
  Statistical analysis of wind speed fluctuation and increments of non-stationary
  atmospheric boundary layer turbulence
}


\begin{center}
  \LARGE
  Statistical analysis of wind speed fluctuation and increments of non-stationary
  atmospheric boundary layer turbulence
\end{center}
\centerline{\large
  T.~Laubrich\textsuperscript{1},  \hfill  
  F.~Ghasemi\textsuperscript{2},  \hfill
  J.~Peinke\textsuperscript{3}, \hfill
  H.~Kantz\textsuperscript{1}
}
\begin{itemize}
\item[\textsuperscript{1}]
         Max-Planck-Institut f\"ur Physik komplexer Systeme, \\
	D-01187 Dresden, Germany\\
	{\tt laubrich@pks.mpg.de}
\item[\textsuperscript{2}]
	Institut f\"ur Theoretische Physik der \\
	Westf\"alischen Wilhelms-Universit\"at M\"unster, \\
    	D-48149 M\"unster, Germany
\item[\textsuperscript{3}]
	Institut f\"ur Physik der Universit\"at Oldenburg,\\
    	D-26111 Oldenburg, Germany 
\end{itemize}

\begin{abstract}
We study the statistics of the horizontal component of atmospheric boundary
layer wind speed. Motivated by its non-stationarity, we investigate which
parameters remain constant or can be regarded as being piece-wise constant and
explain how to estimate them. We will verify the picture of natural atmospheric
boundary layer turbulence to be composed of successively occurring close to
ideal turbulence with different parameters.
 
The first focus is put on the fluctuation of wind speed around its mean
behaviour.  We describe a method estimating the proportionality factor between
the standard deviation of the fluctuation and the mean wind speed and analyse
its time dependence.  The second focus is put on the wind speed increments. We
investigate the increment distribution and use an algorithm based on
superstatistics to quantify the time dependence of the parameters describing
the distribution.

Applying the introduced tools yields a comprehensive description of the wind
speed in the atmospheric boundary layer.
\end{abstract}


\section{Introduction}
\label{S:intro}
  
The statistical analysis of turbulent flow has a long tradition and has
revealed a lot of insight into the properties of turbulence, starting with the
pioneering works of \cite{kolmogorov41c,kolmogorov41a,kolmogorov62}.
However, in the transition regimes between isotropic turbulence as one
idealisation and laminar flow as another idealisation, our knowledge is still
incomplete. This is even more the case when turbulence outside the laboratory
is studied. The air flow in the atmospheric boundary layer (ABL), i.e.\ in the
lowest few 1--2~km of the atmosphere \cite[see][]{wallace06}, is strongly
influenced by surface roughness and hence orography and land use, but even more
by geothermal effects through heating from the ground. Both effects do not only
introduce additional structures into the turbulent flow, but also cause
non-stationarities because these effects depend, e.g.\ on the intensity of solar
radiation and on the direction of the surface wind, which both change much
faster than large scale pressure differences which generate the overall wind
conditions.

In several applications, a better understanding of the statistical properties
of boundary layer turbulence under realistic conditions is essential, in
particular in view of the cost efficient use of wind power. One example is that
more realistic input wind fields than just laminar flows are desired for
numerical simulations of the flow around an obstacle. Another example is the
need of good statistical evidence of extreme wind gusts, their relative
frequency, their spatial extension, and also their temporal correlations for
the estimation of loads on structures and their expected lifetimes.

Motivated by these considerations, we will here discuss the detailed analysis
of boundary layer wind fields.  Moreover, since theoretical concepts such as
that of \cite{CGH90} were developed for an idealised
turbulence, we investigate in how far these results hold true for ABL
turbulence.  We therefore consider time series recordings obtained by a single
anemometer at fixed height above ground.  The wind field at position $\vec r$
and time $t$ is denoted by $\vec u(\vec r,t)$.  The time series is given by the
horizontal component
\begin{equation} \label{E:series}
 	x_n 
	= 
	\sqrt{ 
	    u_x^2\left(\vec r,\frac{n}{\nu}\right) 
	  + u_y^2\left(\vec r,\frac{n}{\nu}\right) 
	}
\end{equation} 
for $n=0,1,2,\dots$ and $\nu$ standing for the measurement frequency.  Making
use of the \cite{taylor38} hypothesis, i.e.\ temporal correlations can
be translated into spatial longitudinal correlations, our analysis aims at a
quantitative characterisation of the statistics of horizontal wind speed
data. 

Throughout our analysis the non-stationarity of the data plays a major role and
we intend to answer the question which parameters remain constant or can be
regarded as being piece-wise constant and how to estimate them.  We draw the
conclusion that natural ABL turbulence is a composition of successively
occurring close to ideal turbulence with different parameters.
 
It is worth mentioning that the same conclusion was drawn in a recent work by
\cite{boettcher07}. However, the authors applied a different
statistical method and concentrated on the wind speed increments.

As for experimental data we study wind speed recordings acquired at 10~m altitude
with a frequency of 8~Hz at the \cite{lammefjord} site. It clearly exemplifies
ABL turbulence. The results of data gathered at 20~m and 30~m above ground do
not differ qualitatively.
 
The plan of the paper is as follows. In the first part we study the fluctuation
of the horizontal wind speed around its mean behaviour.  Empirically, the
standard deviation of the fluctuation grows linearly with the mean wind speed.
We explain a method how to estimate the proportionality factor.  The second
part deals with the statistics of wind speed increments.  The empirical results
are compared to theoretical works which assume the correctness of the
intermittency hypothesis of turbulence 
\cite[see][]{kolmogorov62,obukhov62}.  These works state that the distribution of
short time increments is strongly leptokurtic. The parameters describing this
distribution are in good approximation piece-wise constant. The
superstatistical approach, which the third part of this paper deals with, is
sensitive enough to actually quantify the dynamics of the distribution
parameters. Finally, the last section contains the conclusions.
 

\section{Conditioned Fluctuation Distributions}
\label{S:fluc}

\begin{figure}
	\centerline{
	  \includegraphics{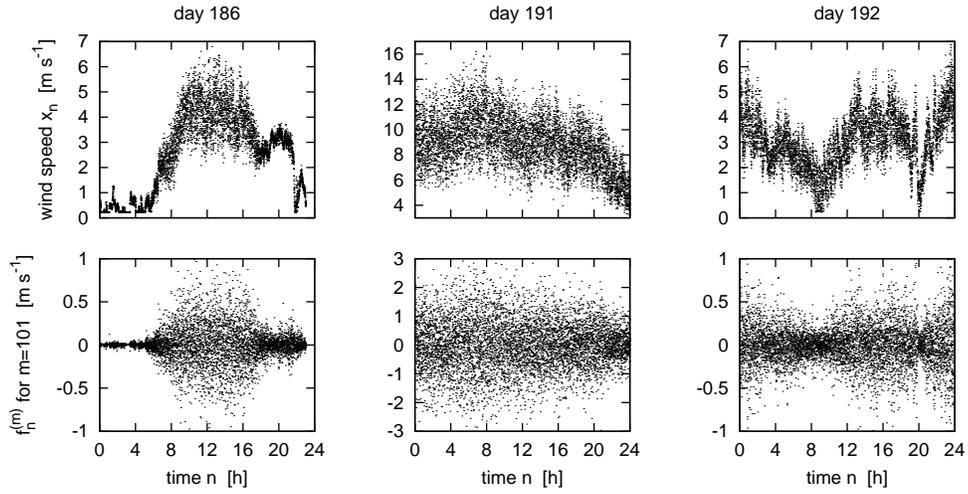}
	}
	\caption{
	  The top row shows the wind speed data for three different days of the
	  \cite{lammefjord} measurement. The second row displays the
	  fluctuation according to~\eqref{E:def:f} for $m=101$ which
	  corresponds to a time window of 12.5~s
	}
	\label{F:f}
\end{figure}

\begin{figure}
	\centerline{
	  \includegraphics{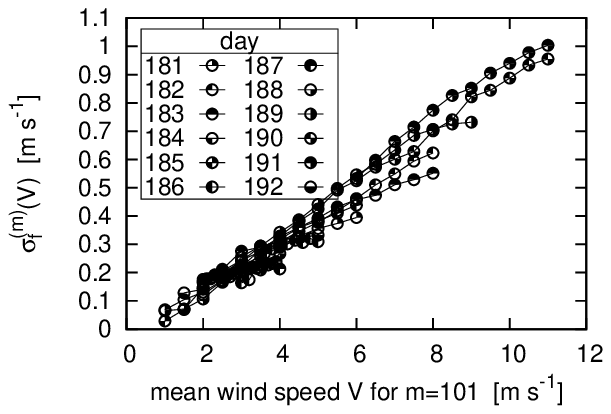}
	  \includegraphics{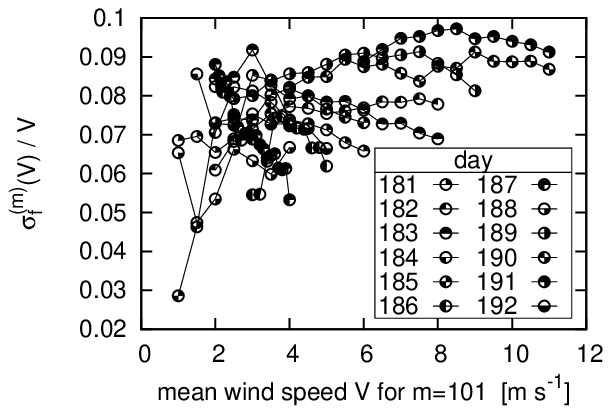}
	}
	\caption{
	  The fluctuation standard deviation $\sigma_f^{(m)}(V)$ (left) and
	  $\sigma_f^{(m)}(V)/V$ (right) as a function of the $m=101$ window
	  mean wind speed $V$ estimated from 24~h recordings at the 
	  \cite{lammefjord} site
	}
	\label{F:propto}
\end{figure}

The first method which we want to give an introduction to analyses the
fluctuation of the wind speed around its window mean over $m=2\tilde m+1$
sample points with $\tilde m=0,1,2,\dots$. In other words, we consider the
fluctuation series
\begin{equation} \label{E:def:f}
 	f^{(m)}_n = x_n - \bar x^{(m)}_n
\end{equation} 
where
\begin{equation} 
 	\bar x^{(m)}_n = \frac{1}{m} \sum_{k=-\tilde m}^{\tilde m} x_{n+k}.
\end{equation} 
Figure~\ref{F:f} (top row) shows three days of measurement at the 
\cite{lammefjord} site illustrating the non-stationarity of ABL wind speed. The
same kind of non-stationarity is inherited in $\bar x^{(m)}_n$ so that the mean
of the fluctuation $f^{(m)}_n$ is (at least nearly) constant with $n$, namely
zero. The second row of the figure displays the fluctuation series
$f_n^{(m)}$ for $m$ chosen exemplarily to be 101. Thus, the fluctuation
corresponds to the wind speed deviation at time $n$ from the 12.5~s window mean
around $n$.  It can be seen that the fluctuation series is centered around zero
and that its volatility becomes larger as the wind speed $x_n$, and hence the
mean wind speed $\bar x^{(m)}_n$, increases. We are interested in the
functional dependence of the volatility of the fluctuation from the mean wind
speed.  Hence, collecting the events 
\begin{equation} 
 	n\in\NmV = \{n:\bar x^{(m)}_n = V\}
\end{equation} 
allows us to estimate numerically the variance of the set
$\left\{f^{(m)}_n:n\in\NmV\right\}$ for each 24~h
sample individually. This variance can be identified with the 
conditioned volatility  $\sigma_f^{(m)}(V)^2$ under the assumption that the
conditioned variance $\Var{f^{(m)}_n|V}$ does not depend on $n\in\NmV$.
Figure~\ref{F:propto} shows the empirical result: The proportionality
$\sigma_f^{(m)}(V)\propto V$ can be verified for each 24~h data individually.
Figure~\ref{F:propto} (right) reveals that the proportionality becomes better
as $V$ becomes larger. It is shown in figure~\ref{F:f} that the wind speed can
be low during the night hours where the wind is less turbulent causing
$\sigma_f^{(m)}(V)/V$ to be comparably small.
 
The proportionality factor differs slightly from day to day. 
Consequently, splitting the set \NmV\ into two disjoint subsets ${\cal N}^{(m)}_1(V)$
and ${\cal N}^{(m)}_2(V)$ whose union is \NmV, ABL data show that
the variance of the set $\{f^{(m)}_n: n\in{\cal N}^{(m)}_1(V)\}$ 
might not coincide with the variance of 
$\{f^{(m)}_n:n\in{\cal N}^{(m)}_2(V)\}$. 
We therefore need to refine the method.

\begin{figure}
	\centerline{
	  \includegraphics{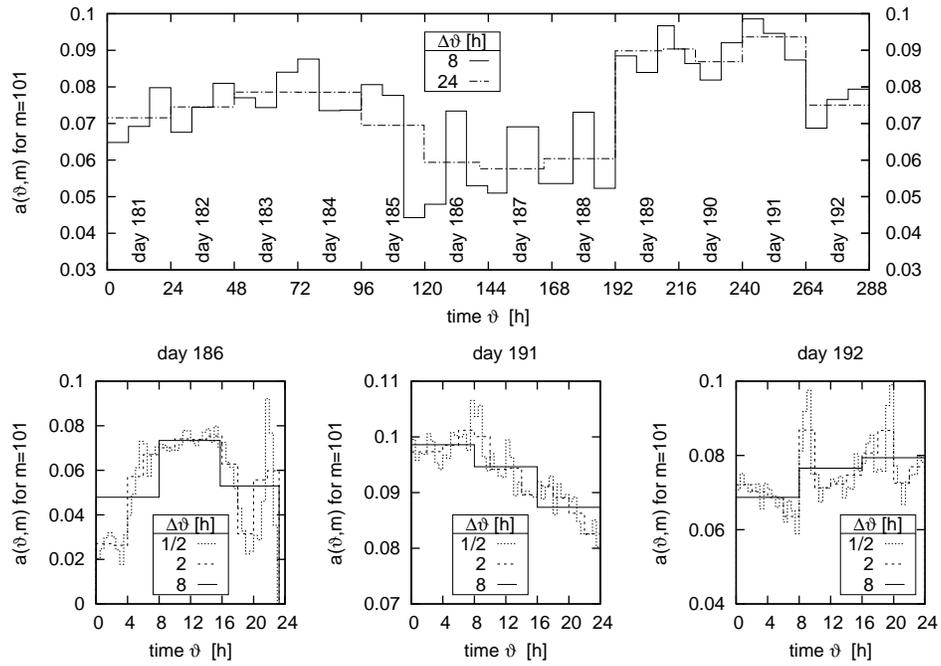}
	}
	\caption{
	  Standard deviation of the normalised fluctuation with $m=101$ at the
	  \cite{lammefjord} site estimated in successive periods of
	  lengths $\Delta\vartheta$. The top panel shows the ``long-term''
	  behaviour over eleven days of measurement ($\Delta\vartheta=8, 24\,{\rm
	  h}$) whereas the lower three panels
	  depict three 24~h samples individually ($\Delta\vartheta=1/2, 2, 8\,{\rm
	  h}$)
	}
	\label{F:a}
\end{figure}

From a meteorological point of view, the considered variances are mainly
determined by the stratification and thus the Richardson number.  The ABL wind
field is strongly influenced by geothermal effects changing with time
(day/night cycle, clouds etc.).  We should therefore expect 
\begin{equation} 
 	\sqrt{\Var{f^{(m)}_n}} = a_n(m) \bar x^{(m)}_n
\end{equation} 
with a time dependent $a_n(m)$. We assume that the proportionality factor
changes on a larger time scale than the fluctuation so that it
can be approximated by being piece-wise constant over $\Delta\vartheta$ time
steps. We therefore assume
\begin{equation} 
 	a_n(m) \approx a(\vartheta,m)
\end{equation} 
for $\vartheta-\frac{\Delta\vartheta}{2}\le n <\vartheta+\frac{\Delta\vartheta}{2}$
where $\vartheta$ represents the time of the day.  
Defining
\begin{equation} 
 	\NmtV 
	= 
	\left\{
		\vartheta-\frac{\Delta\vartheta}{2}
		\le n <
		\vartheta+\frac{\Delta\vartheta}{2}
		: \bar x^{(m)}_n = V
	\right\}
\end{equation} 
and assuming that the volatility of the fluctuation at time $n\in\NmtV$
does not depend on $n$, the normalised fluctuation
\begin{equation} \label{E:def:g}
 	g^{(m)}_n 
	= 
	\frac{f^{(m)}_n}{\bar x^{(m)}_n},
\end{equation} 
with $\bar x^{(m)}_n\ne0$
has a time-independent volatility for $n\in\NmtV$. 
We denote the mentioned volatilities by
$\sigma^{(m)}_f(\vartheta,V)^2$ and $\sigma^{(m)}_g(\vartheta,V)^2$, respectively,
and write
\begin{equation} 
 	\sigma^{(m)}_g(\vartheta,V) = \frac1V \sigma^{(m)}_f(\vartheta,V)
\end{equation} 
so that
\begin{equation} \label{E:multi}
 	\sigma_f^{(m)}(\vartheta,V) = a(\vartheta,m) V 
	\iff 
	\sigma_g^{(m)}(\vartheta,V) = a(\vartheta,m).
\end{equation} 
In other words, the standard deviation of the fluctuation grows linearly
with $V$ if the standard deviation of the normalised fluctuation does not
depend on $V$. The latter is equivalent to saying that the volatility of
$g^{(m)}_n$ remains constant
for $\vartheta-\Delta\vartheta/2\le n <\vartheta+\Delta\vartheta/2$.
Under this assumption we can estimate the proportionality factor
$a(\vartheta,m)$ by computing
the standard deviation of the set $\left\{g^{(m)}_n:n\in\NmtV\right\}$.
Figure~\ref{F:a} depicts the so obtained proportionality factors over
a period of eleven days of measurement.
The top panel contains $a(\vartheta,m=101)$ as a step 
function of time $\vartheta$ with 
$\Delta \vartheta=8$ and 24~h. The lower three panels depict the proportionality
factor approximated by a step function with 1/2, 2, and 8~h plateaus
exemplarily for the days 186, 191, and 192 of the measurement. It can be seen
that $a(\vartheta,m)$ changes with time leading to the question whether its
approximation of being constant over a period is suitable.

\begin{figure}
	\centerline{
	  \includegraphics{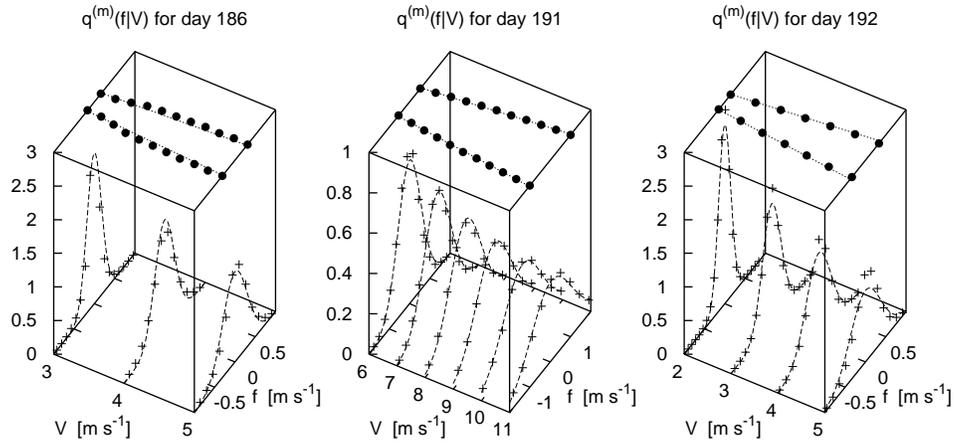}
	}
	\caption{
	  Estimated distribution of the fluctuation conditioned on mean wind speed
	  $V$ over $m=101$ time steps for the wind data acquired at the 
	  \cite{lammefjord} site.  The crosses symbolise
	  the estimation whereas the dashed lines represent Gaussian distributions
	  given in~\eqref{E:qf:gauss}
	  with $a(\vartheta,m)$ being estimated by the 24~h standard deviation
	  of the normalised fluctuation, see~\eqref{E:est:a24h} and
	  figure~\ref{F:a} (top panel).  
	  The standard deviation of the fluctuation is again plotted on top of
	  the boxes. The black
	  dots represent the numerical estimation whereas the dotted line
	  corresponds to a line through the origin with slope $a(\vartheta,m)$
	}
	\label{F:chist}
\end{figure}	
We check if treating $a(\vartheta,m)$ as remaining constant over 24~h is
a good approximation by estimating the histograms
of the fluctuation conditioned on a window mean wind speed $V$ and
comparing it with symmetric Gaussian distributions \footnote{Under the
assumption that $\E{x_n|V}=V$, the mean of the fluctuation vanishes, i.e.\
$\E{f^{(m)}_n|V} = 0$.} whose standard deviation is proportional to V. The
proportionality factor is set equal to the estimated standard deviation of the
set $\left\{g^{(m)}_n:n\in\NmtV\right\}$.  That is, computing
\begin{equation} 
 	q^{(m)}(f|V) = \mean{\delta(f-f^{(m)}_n)}_{n\in\NmtV}
\end{equation} 
and comparing it with
\begin{equation} \label{E:qf:gauss}
 	\hat q^{(m)}(f|V) 
	=
 	\frac1{\sqrt{2{\upi}}a(\vartheta,m)V}{\rm e}^{-\frac{f^2}{2a(\vartheta,m)^2V^2}}
\end{equation} 
where
\begin{equation} \label{E:est:a24h}
 	a(\vartheta,m) 
	= 
	\sqrt{
	  \VarS{g^{(m)}_n:
	  \vartheta-\frac{\Delta\vartheta}{2}\le n < \vartheta+\frac{\Delta\vartheta}{2}}
	}
\end{equation} 
for $\Delta\vartheta=24~{\rm h}$. Figure~\ref{F:chist} shows the
estimated histograms for $m=101$ and $\vartheta$ representing day 181, 191, and
192 of the Lammefjord measurement. It can be seen that they are
in good agreement with the above mentioned Gaussian distributions 
verifying
\begin{equation} 
 	\sigma_f^{(m)}(V) \propto V.
\end{equation} 
On top of the boxes the conditioned standard deviation of the fluctuation
is drawn vs.\ $V$. The dotted line represents a proportional 
dependence with an $a(\vartheta,m)$ estimated via~\eqref{E:est:a24h}.
It can be concluded that the 24~h estimation yields reasonable results.

As a side remark, the estimated constant $a(\vartheta,m)$
is also referred to as the turbulence
intensity (\ti)\ as defined in \cite{we_handbook} examined over the period 
$T=m/\nu$ around the time $\vartheta$.  It is customary in turbulence
research to decompose the wind speed, in our case the horizontal component,
during a period of time $T$ as
\begin{equation} 
 	u(\vec r,t) = U(\vec r) + u'(\vec r,t)
\end{equation} 
where $U(\vec r)$ is the average value of $u(\vec r,t)$ over the period $T$
which is typically chosen to be 10~min or 1~h.  The
\ti\ is defined as the root mean
square of $u'(\vec r,t)$ over the period $T$ in units of $U(\vec r)$ and it
states the percentage of the mean flow which are represented by the velocity
fluctuation.  The assumed linear relation between the standard deviation of the
fluctuation and the mean wind speed is in accordance with this interpretation.
The \ti\ describes the strength of the instantaneous turbulence at time
$\vartheta$ and does for instance depend on the weather situation.  Hence, it
can vary from one measurement period to another measurement period.
 
The proportionality $\sigma^{(m)}_f(V)\propto V$ is not a generic property
of a random process.  For instance, the variance of the fluctuation around
$V$ in a white noise (wn) process $\xi_n$ does not depend on $V$, i.e.\ 
$\sigma^{(m)}_{f;{\rm wn}}(V)={\rm const}$.  
The independence between $\xi_n$ and $\xi_{n+s}$ for $s\ne0$ 
allows to conclude that $\E{\xi_n|V}=\E{\xi_{n+s}|V}=V$ and
$\E{\xi_n^2|V}=\E{\xi_{n+s}^2|V}$ for $|s|<\tilde m$. 
The conditioned variance of the fluctuation
is equivalent to the expectation of the biased variance estimation of
the set $\{\xi_{n-\tilde m}, \dots, \xi_{n+\tilde m}\}$ with sample mean $V$
and therefore independent from $V$:
\begin{align} 
	\E{(\xi_n - V)^2|V}
	&=
	\frac1m \sum_{s=-\tilde m}^{\tilde m}\E{(\xi_{n+s} - V)^2|V}
	\notag \\
	&=
	\E{\frac1m \sum_{s=-\tilde m}^{\tilde m}(\xi_{n+s} - V)^2\Big|V}
	\notag \\
	&=
	\frac{m-1}{m}.
\end{align} 

In general, any stationary process $y_n$ with mean $\E{y_n}$ does not show a
proportionality between the standard deviation of the fluctuation and the
window mean. This is because the variance of the numerically estimated
series $\bar y^{(m)}_n$ is nearly zero for sufficiently large $m$. 
The mean of $\bar y^{(m)}_n$ coincides with the global mean 
$\E{y_n}$. In other
words, $\bar y^{(m)}_n\sim{\rm const}$ so that the variance of the fluctuation
conditioned on $\bar y^{(m)}_n=V$ can only be evaluated if $V=\E{y_n}$ making
the condition in fact redundant.  The distribution of the fluctuation coincides
with the distribution of $y_n$ shifted by $-\E{y_n}$.  

\begin{figure}
	\centerline{
	  \includegraphics{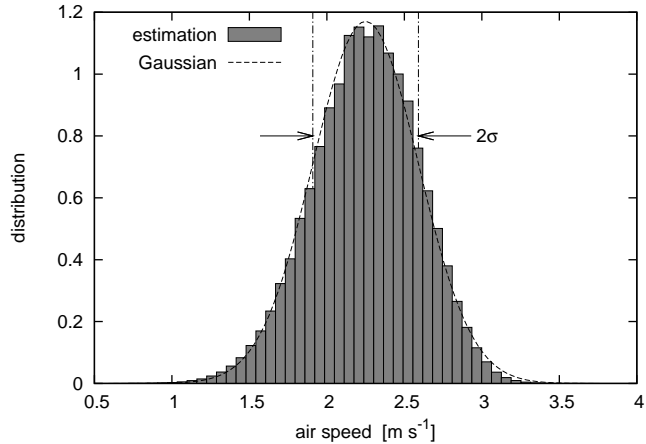}
	}
	\caption{
	  The distribution of the air speed measured in the free jet
	  experiment \cite[see][]{freejet}. The dashed line corresponds to 
	  a Gaussian curve with mean 2.25~$\rm m\,s^{-1}$ and standard
	  deviation 0.341~$\rm m\,s^{-1}$
	}
	\label{F:free:hist}
\end{figure}	
\cite{freejet} measured the air speed in an
air into air round free jet experiment. The acquired data series is 
an example of stationary laboratory turbulence. 
Figure~\ref{F:free:hist} shows the distribution of the
air speed in the free jet experiment which is in good agreement with a
Gaussian shaped curve. In other words, the shown histogram corresponds to 
one slice of the three-dimensional plots in figure~\ref{F:chist}.

As a conclusion, analysing a period $\vartheta-\Delta\vartheta/2\le
n<\vartheta+\Delta\vartheta/2$ of ABL wind speed data reveals that
the distribution of the fluctuation $f^{(m)}_n$ 
around the mean speed $\bar x^{(m)}_n=V$ is well described
by a symmetric Gaussian distribution with a standard deviation being
proportional to $V$. The proportionality factor $a(\vartheta,m)$ can be
estimated by evaluating the standard deviation of the normalised fluctuation
$f^{(m)}_n/\bar x^{(m)}_n$ over the period of interest.
 
In a further study, we propose a stochastic process which
can be used to generate a time series having the same fluctuation behaviour as
ABL wind speed data. The work is still in progress.
 
Additionally, a conclusion about wind speed increments
$x_{s;n}=x_{n+s}-x_n$ for sufficient large $s$ can be drawn. If
$x_{n+s}$ and $x_n$ can be treated as being independent from each other and
choosing $m$ such that $\bar x^{(m)}_n\approx \bar x^{(m)}_{n+s}$,
the increment $x_{s;n}$ is of a symmetric normal distribution with a standard
deviation being proportional to $\bar x^{(m)}_n$.
Taking the statistics of $x_{s;n}$ over a much larger time period than the
time scale on which $\bar x^{(m)}_n$ changes, e.g.\ 24~h, the increment
distribution $p_s(x_s)$ corresponds to a superposition of symmetric Gaussians 
with different volatilities. The variance of the volatilities is determined by
the variance of $\bar x^{(m)}_n$. It (nearly) vanishes if
$\bar x^{(m)}_n\sim{\rm const}$ in time making $p_s(x_s)$ coincide with
a Gaussian distribution.  Otherwise, $p_s(x_s)$ is fat tailed.

The following two sections intend to analyse the increment series,
its statistics, and its limit for large increment lengths.

 
\section{Increment Statistics}
\label{S:inc}
 

This section is dedicated to the statistics of wind speed increments. The
increment series is defined as
\begin{equation} \label{E:def:incseries}
 	x_{s;n} = x_{n+s}-x_n
\end{equation} 
with $s$ denoting the time over which the increment is measured---the increment
length.


\cite{CGH90} deduced  an analytical expression for the
marginal increment distribution $p_s(x_s)$ from the assumption of a
log-normally distributed energy transfer rate in turbulence.  Castaing's
hypothesis assumes that the increment series in a small time window is of a
Gaussian distribution
\begin{equation} 
 	p(x_s|\beta) 
		= \sqrt{\frac{\beta}{2\upi}}{\rm e}^{-\frac{\beta}{2}x_s^2}.
\end{equation} 
In each window the parameter $\beta$ can be regarded as being constant. However,
it varies between the windows according to the log-normal
distribution 
\begin{equation} \label{E:lognormalfbeta}
	f_s(\beta) 
	= \frac{1}{\sqrt{2\upi}\lambda_s\beta}
	   {\rm e}^{-\frac12
           \left(\frac{1}{\lambda_s}\ln\frac{\beta}{\beta_s}\right)^2
		 }.
\end{equation} 
As a result, the increment distribution of ABL wind speed is given by
\begin{align}	
 	p_s(x_s) 
	&= \int_{0}^{\infty} \D \beta\, f_s(\beta)\, p(x_s|\beta) \notag \\
	&= \frac{1}{2\upi\lambda_s}
	   \int_0^\infty \D \beta  
	          \frac{1}{\sqrt{\beta}}
		      {\rm e}^{
		      	-\frac12 \left( \frac{1}{\lambda_s} 
				                \ln \frac{\beta}{\beta_s} \right)^2
			    -\frac12 \beta x_s^2
		      }.
	\label{E:castaing}
\end{align}	
It is symmetric, leptokurtic, and described by
two positive parameters $\beta_s$ and $\lambda_s$ which are called 
position and shape parameter, respectively. The latter is directly related
to the kurtosis of the increment series by
\begin{equation} 
	k_s = \frac{\mean{x_s^4}}{\mean{x_s^2}^2} = 3{\rm e}^{\lambda_s^2} > 3
\end{equation} 
so that the shape parameter can be estimated using
\begin{equation} \label{E:est:lambda2}
 	\lambda_s^2 = \ln \frac{k_s}{3}.
\end{equation} 
The position parameter of the 
increment process can be estimated via the variance and kurtosis by
\begin{equation} \label{E:est:beta}
 	\beta_s = \frac{1}{\sigma_s^2}\sqrt{\frac{k_s}{3}}.
\end{equation} 
In the limit of $\lambda^2_s\to0$ the volatility distribution $f_s(\beta)$
in~\eqref{E:lognormalfbeta} turns into
\begin{equation} 
 	f_s(\beta) 
	\overset{\lambda^2_s\to0}{\longrightarrow} 
	\delta(\beta-\beta_s)
\end{equation} 
making the increment distribution coincide with a Gaussian distribution
with variance $\sigma_s^2=1/\beta_s$ and kurtosis $k_s = 3$:
\begin{equation} 
	p_s(x_s)
	\overset{\lambda^2_s\to0}{\longrightarrow} 
	\sqrt{\frac{\beta_s}{2\upi}}{\rm e}^{-\frac{\beta_s}{2}x_s^2}.
\end{equation} 

\begin{figure}
	\centerline{
	  \includegraphics{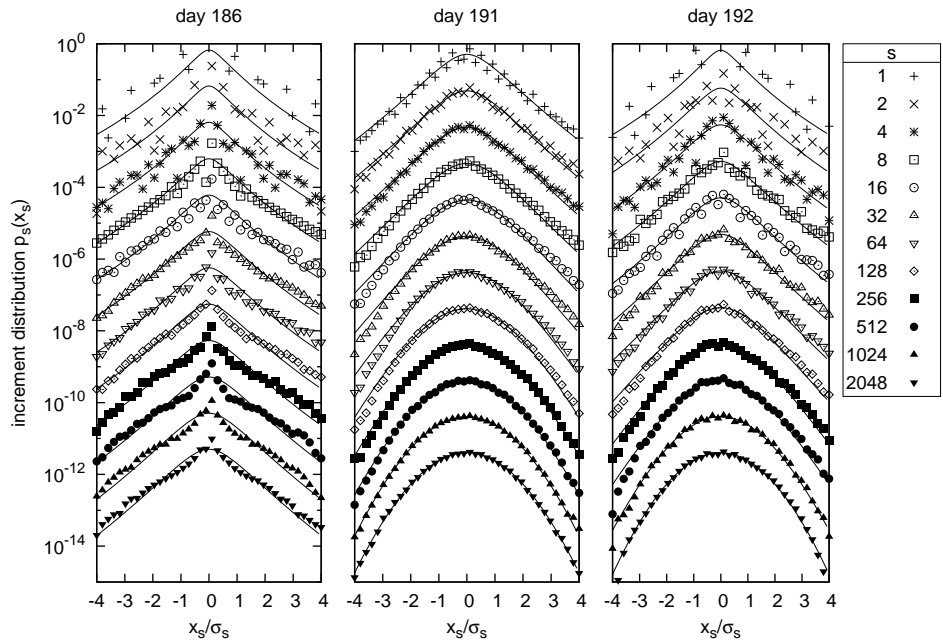}
	}
	\caption{
	  Increment histograms $p_s(x_s)$ for day 186
	  (left), day 191 (centre), and day 192 (right) of the \cite{lammefjord}
	  measurement.  The normalised histograms are drawn with points
	  whereas the solid lines represent a Castaing distribution with the same
	  variance and kurtosis as the increment series. The histograms are
	  shifted and drawn in a semi-logarithmic plot for better visibility
	}
	\label{F:inc:hi}
\end{figure}	
Figure~\ref{F:inc:hi} depicts the increment distribution $p_s(x_s)$ for a variety
of $s$ and for three different days (day 186, 191, and 192) of the
\cite{lammefjord} measurement.  The fitted Castaing distributions
according to~\eqref{E:est:lambda2} and \eqref{E:est:beta} are drawn with
solid lines. It can be seen that the empirical histograms are in good agreement
with Castaing distributions.  Furthermore, the histograms for the data acquired
at the days 191 and 192 approach a Gaussian distribution as $s$ becomes larger.

\begin{figure}
	\centerline{
	  \includegraphics{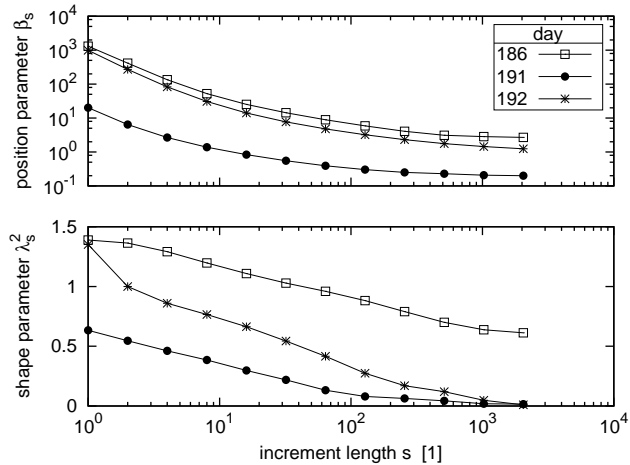}
	}
	\caption{
	  The position parameter $\beta_s$ and shape parameter $\lambda^2_s$ as a
	  function of $s$ for the days 186, 191, and 192 of the 
	  \cite{lammefjord} measurement are plotted in the upper and lower panel,
	  respectively
	}
	\label{F:inc:hipar}
\end{figure}	
The intermittency hypothesis and the conclusion drawn in Sec.~\ref{S:fluc}
imply that the shape parameter $\lambda^2_s$ decreases as $s$ gets larger. In
case of (nearly) stationary turbulence it converges to zero, i.e.\
\begin{equation} \label{E:lambda2to0}
	\lambda^2_s\overset{s\to\infty}{\longrightarrow}0.
\end{equation} 
\cite{boettcher02} obtained results from laboratory and ABL turbulence 
showing~\eqref{E:lambda2to0}.
Consequently, with increasing $s$ both the kurtosis $k_s$ and inverse variance
$1/\sigma_s^2$ \footnote{The variance becomes larger due to $\sigma^2_s =
2\sigma^2(1-\gamma(s))$ where $\sigma^2$ and $\gamma(\cdot)$ denote the
variance and auto correlation function of the series $x_n$, respectively.}
decrease causing the position parameter to decrease, too.  
Figure~\ref{F:inc:hipar} depicts $\beta_s$ vs.\ $s$ and $\lambda^2_s$ vs.\ $s$ for
the increment distributions of the Lammefjord ABL wind speed data.  
It can be seen that the shape parameter $\lambda^2_s$
of the days 191 and 192 goes to zero as $s$ becomes larger whereas it does not
reach zero for the day 186 data. In other words, the data acquired at day 186
have leptokurtic increment distributions at large increment lengths. 

It is worth mentioning that for laboratory turbulence the empirical increment
histograms are in perfect agreement with the hypothetical Castaing shaped
curves \cite[see][figure~8]{freejet}.  Additionally, laboratory turbulence data
yield a shape parameter which decreases with $s$ according to a power law and
approaches zero for sufficiently large $s$ \cite[see][figure~9]{freejet}.
 
As a summary,
this simple approach showed an agreement between the empirical
histograms and the hypothetical distributions. 
The effect of increasing
Gaussianity and hence decreasing shape parameter for increasing $s$ 
can be seen and is well supported by the intermittency hypothesis of idealised 
turbulence.  

Some measurements show a fat tailed increment distribution even for large
increment lengths $s$. According to \cite{boettcher07} and the conclusion drawn
in Sec.~\ref{S:fluc}, a fluctuating mean wind speed might be the reason for
this behaviour.  It is however the drawback of this technique not being
suitable to verify a connection between this kind of non-stationarity and the
shape of increment distributions.
 

\section{Superstatistics}
\label{S:super}
 

We scrutinise the increment series a little further in order to understand
the different increment statistics behaviour of ABL wind.
The method which this section has its focus on aims to 
validate the Castaing hypothesis~\eqref{E:castaing} in another way: by
computing the, a priori unknown, distribution $f_s(\beta)$ and comparing it
with a log-normal distribution.
 

The concept of superposing two statistics was generalised by 
\cite{BC03} and called ``superstatistics''.  It describes a driven
non-equilibrium system  of an intensive parameter $\beta$, which  in our case is the
inverse volatility, fluctuating on a large spatio-temporal scale $T$
whereas the system itself changes on a short spatio-temporal scale $\tau \ll
T$.  

\begin{figure}
	\centerline{
	  \includegraphics{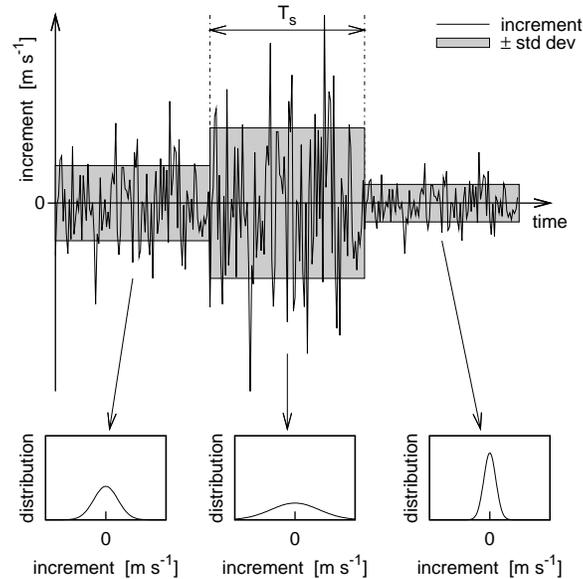}
	}
	\caption{
	  Schematic sketch of the superstatistical approach. The increment series
	  is considered to consist of successive normally distributed segments
	  of length $T_s$. Within such a period the mean is considered to be zero
	  and the standard deviation (std dev) approximated by the estimation over
	  the interval
	}
	\label{F:super:sample}
\end{figure}	
Based on this idea \cite{beck05_b} and \cite{queiros07} proposed
an algorithm to approximate the volatility by being piece-wise constant,
namely $1/\beta$, and treat
the increment series as being of a Gaussian distribution in each segment.
Figure~\ref{F:super:sample} sketches schematically the approximation. 
 
We use this algorithm to estimate the statistics of $\beta$ from the increment
series $(x_{s;n})_{n=0}^{N-1}$ of ABL wind speed data.  The essential step is
to find the large time scale $T_s$ and estimate $\beta$ in each segment.
 
\begin{figure}
	\centerline{
	  \includegraphics{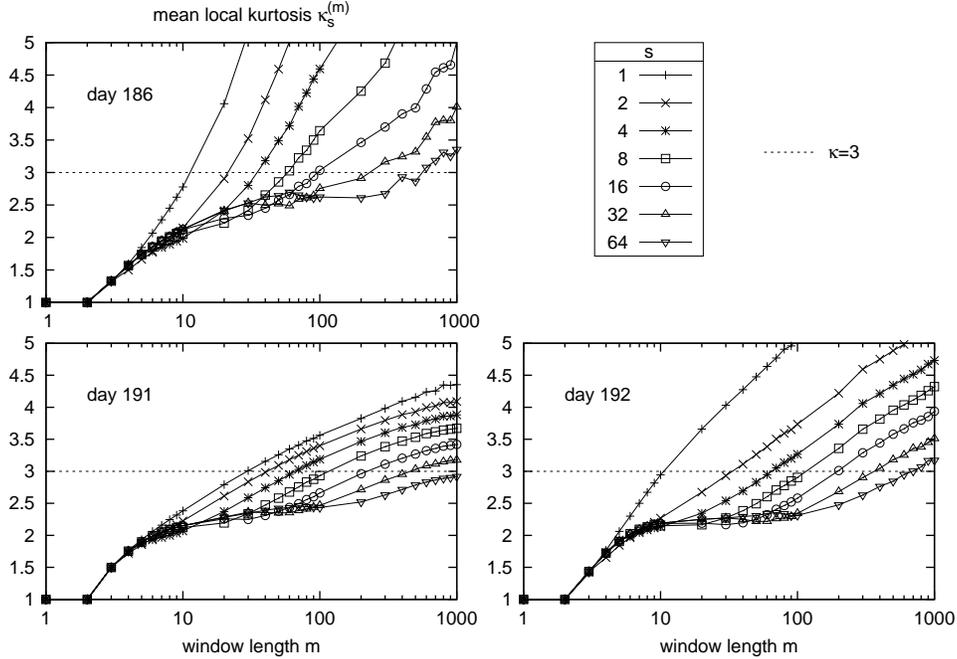}
	}
	\caption{
	  Mean window kurtosis as function of the window size $m$ for
	  different increment length $s$ of the ABL wind speed measurement at the
	  \cite{lammefjord} site. The dashed line corresponds to
	  $\kappa=3$. The intersection $\kappa_s^{(m)}=3$
	  is an estimation for the large time scale shown in 
	  figure~\ref{F:super:ts}
	}
	\label{F:super:kurtosis}
\end{figure}	
The large time scale can be identified with the scale on which the increment series is
of a normal distribution. Hence, being a measure for Gaussianity, the sample kurtosis
of a population ${\cal A}$ is defined as
\begin{equation} 
 	\KurtO{{\cal A}} 
	= 
    |{\cal A}|\times
	\frac{\sum_{a\in{\cal A}} (a-\bar a)^4}
	     {\left(\sum_{a\in{\cal A}} (a-\bar a)^2\right)^2}
\end{equation} 
with $\bar a=\frac{1}{|{\cal A}|}\sum_{a\in{\cal A}} a$ denoting the sample
average and $|{\cal A}|$ denoting the sample size. If the population stems from
a Gaussian distributed population, the kurtosis equals to three if the 
population size is reasonably large. The above defined sample kurtosis is biased 
with respect to the sample size. If ${\cal A}$ consists of only one or two 
elements, i.e.\ ${\cal A}=\{a\}$ or ${\cal A}=\{a_1,a_2\}$, the sample
kurtosis $\KurtO{{\cal A}}=1$. In order to find the large time scale,
the increment series $(x_{s;n})_{n=0}^{N-1}$
is split into $N_m$ sub-series: ${\rm int}\left[N/m\right]$
sub-series of size $m$ and if necessary one remaining sub-series of size
less than $m$. For each sub-series the sample
kurtosis is computed. The average of the $N_m$ sub-series kurtosis is taken
as measure of sub-series Gaussianity. Changing $m$ means changing the length of
the sub-series so that we arrive at the mean sub-series kurtosis
written as
\begin{equation} \label{E:super:kurt}
 	\kappa^{(m)}_{s} 
	= \mean{ \KurtS{x_{s;km},\dots,x_{s;(k+1)m-1}} }_{k=0,\dots,N_m-1}
\end{equation}
being a function of the window length $m$. As mentioned above,
$\kappa^{(m)}_s=1$ for $m\le2$ . As $m$ gets larger $N_m$ tends to one. This
results in a $\kappa^{(m)}_s$ being equal to the sample kurtosis of the whole
increment series whose kurtosis is larger than three because we know from the
increment statistics that the increment series is of a leptokurtic
distribution. Hence, somewhere between $1<m<N$ there is a value $m$ where
$\kappa^{(m=T_s)}_{s}$ is closest to three,
\begin{equation} 
 	\kappa^{(m=T_s)}_{s} = 3,
\end{equation} 
which, according to \cite{beck05_b} and \cite{queiros07}, is taken as an estimation for
the large time scale $T_s$.  Figure~\ref{F:super:kurtosis} displays the mean
window kurtosis as function of the
window size $m$ for different increment lengths $s$ for the ABL wind speed
measurement at the \cite{lammefjord} site. It can be seen that it
increases with increasing $m$ and has an intersection with three.
 
As $T_s$ is supposed to be large there should be no problem with using 
the biased kurtosis estimator defined above.
Using an unbiased estimator would give curves like
$\kappa^{(m)}_s\approx 3$ for $m<T_s$ and $\kappa^{(m)}_s>3$ for
$m>T_s$. From a statistical point of view it is more involved to find the
point $T_s$ that way making the algorithm unnecessarily more complicated.
 
Knowing the length $T_s$ of the sub-series such that they show in average Gaussian
behaviour, leads to the question whether the relaxation time of the process
is small compared to $T_s$. It is reflected by the short time scale
$\tau_s$ being estimated by the decay of the auto correlation function
$\gamma_{s;t}$:
\begin{equation} \label{E:def:tau}
 	\tau_s = \min_{t=1,...,N-1}\{t:\gamma_{s;t} \le {\rm e}^{-1}\}.
\end{equation} 
If $\tau_s$ is much smaller than $T_s$ the existence of two separated time
scales is justified so that it is suitable to estimate the the variance in each
sub-series of length $T_s$ by its (unbiased) sample variance and identify its
inverse as 
\begin{equation} \label{E:def:beta}
 	\beta_k 
	= 
	\frac{1}{\VarS{x_{s;k T_s},\dots,x_{s;(k+1)T_s-1}}}.
\end{equation}
Its estimated distribution reads
\begin{equation} \label{E:def:fbeta}
 	f_s(\beta) = \mean{ \delta(\beta-\beta_k) }_k
\end{equation} 
which, hypothetically, has the shape of a log-normal distribution.
Thus, it is more convenient to consider the variables
\begin{equation} \label{E:def:lambda}
  \Lambda_k = \ln\beta_k
\end{equation} 
and their estimated distribution 
\begin{equation}
 	h_s(\Lambda) 
 	= \mean{\delta(\Lambda-\Lambda_k)}_k.
\end{equation} 
The hypothesis reads
\begin{equation} \label{E:super:hl}
 	H_0:\quad
	h_s(\Lambda) 
		= \frac{1}{\sqrt{2\upi}\lambda_s} 
		  {\rm e}^{-\frac12 
		  	\left(\frac{\Lambda-\Lambda_s}{\lambda_s}\right)^2}
\end{equation} 
with $\Lambda_s=\ln\beta_s$ turning~\eqref{E:castaing} into 
\begin{equation} \label{E:castaing2}
 	p_s(x_s) = \frac{1}{2\upi\lambda_s}\int_{-\infty}^{\infty} \D\Lambda\,
 		{\rm e}^{
		  	-\frac12 \left\{
		  		\left(\frac{\Lambda-\Lambda_s}{\lambda_s}\right)^2
		   		-\Lambda+x_s^2{\rm e}^{\Lambda}
			\right\}
		}.
\end{equation} 
This approach provides an alternative way to fit the Castaing parameters by
\begin{equation} 
 	\ln\beta_s = \Lambda_s = \MeanS{\Lambda_0,\dots,\Lambda_{N_{\Lambda;s}-1}}
	\label{E:super:fit1}
\end{equation} 
and
\begin{equation} 
 	\lambda_s^2 = \VarS{\Lambda_0,\dots,\Lambda_{N_{\Lambda;s}-1}}.
	\label{E:super:fit2}
\end{equation} 

\begin{figure}
	\centerline{\includegraphics{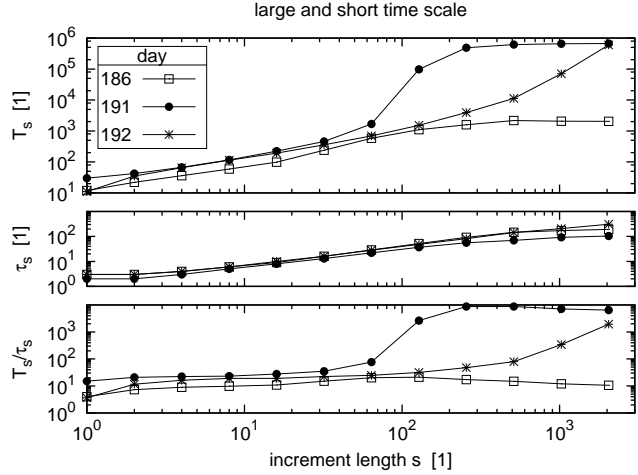}}
	\caption{The top and central panel show the estimated
	         large and short time scales, 
	         $T_s$ and $\tau_s$ in time steps, 
			 respectively, for three 24~h data acquired at
			 the \cite{lammefjord} site (day 186, 191, and 192). 
			 A day consists of 691200 data points
			 being the upper limit for the estimation of $T_s$. The central
			 panel displays the short time scale $\tau_s$ which is estimated by
			 the decay of the auto correlation function for small lags.
			 The ratio $T_s/\tau_s$ is
			 depicted in the lowest panel
	}
	\label{F:super:ts}
\end{figure}
Figure~\ref{F:super:ts} shows
the estimated large and short time scale of wind speed increment data
acquired at the \cite{lammefjord} site as a function of $s$ in its top and
central panel, respectively.  $T_s$, whose estimation is bounded from above by
the number of points which the time
series consists of, generally increases with increasing $s$.  It describes the
scale on which the increment process is of a normal distribution and is hence
a measure for Gaussianity.  Therefore, an increasing large time scale is in
full agreement with the approach to Gaussianity as the increment length gets
larger.  Additionally, the comparably slow approach to Gaussianity of the data
gathered at day 186 can be recovered, cf.\ figure~\ref{F:inc:hipar}.
The plot in the lowest panel of figure~\ref{F:super:ts} verifies the existence of
two separated time scales.  It visualises the ratio $T_s/\tau_s$ being
of the order of magnitude of 10 or larger.

\begin{figure}
	\centerline{
	  \includegraphics{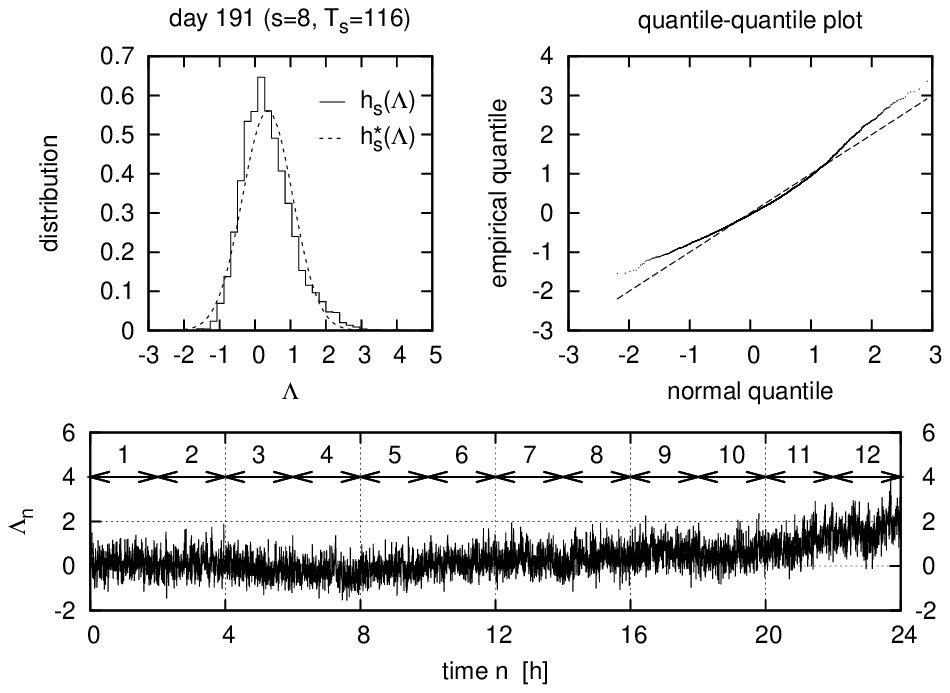}
	}
	\caption{
	  The top left panel displays the estimated 24~h $\Lambda$-distribution
	  $h_s(\Lambda)$ for $s=8$ of the \cite{lammefjord} measurement day 191.
	  $h_s^*(\Lambda)$ denotes a Gaussian distribution with mean
	  $\mean{\Lambda}$ and variance $\mean{(\Lambda-\mean{\Lambda})^2}$.
	  The quantile-quantile plot is shown in the upper right panel.  The bottom
	  panel displays the $\Lambda$-series for that day. The numbers 1 to 12
	  illustrate the twelve 2~h sub-samples which are individually investigated
	  with respect to superstatistics
	}
	\label{F:super:day191:24h}
\end{figure}

The series $(\Lambda_k)_{k=0}^{N_{\Lambda;s}-1}$ for the Lammefjord day 191 data is
computed according to~\eqref{E:def:beta} and \eqref{E:def:lambda}.
Figure~\ref{F:super:day191:24h} shows exemplarily that the estimated distribution
$h_s(\Lambda)$ for $s=8$ is close to Gaussian. But it has a systematic and
statistically significant deviation. In fact, it has a positive skewness.  The
non-Gaussianity is underlined by the quantile-quantile plot in the top right
panel of this figure. However, the increment distribution is in good agreement
with a Castaing shaped distribution, cf.\ figure~\ref{F:inc:hi}.  This allows the
statement that the superstatistical approach is sensitive enough to discover
small deviations from Castaing's hypothesis. 
\begin{figure}
	\centerline{
	  \includegraphics{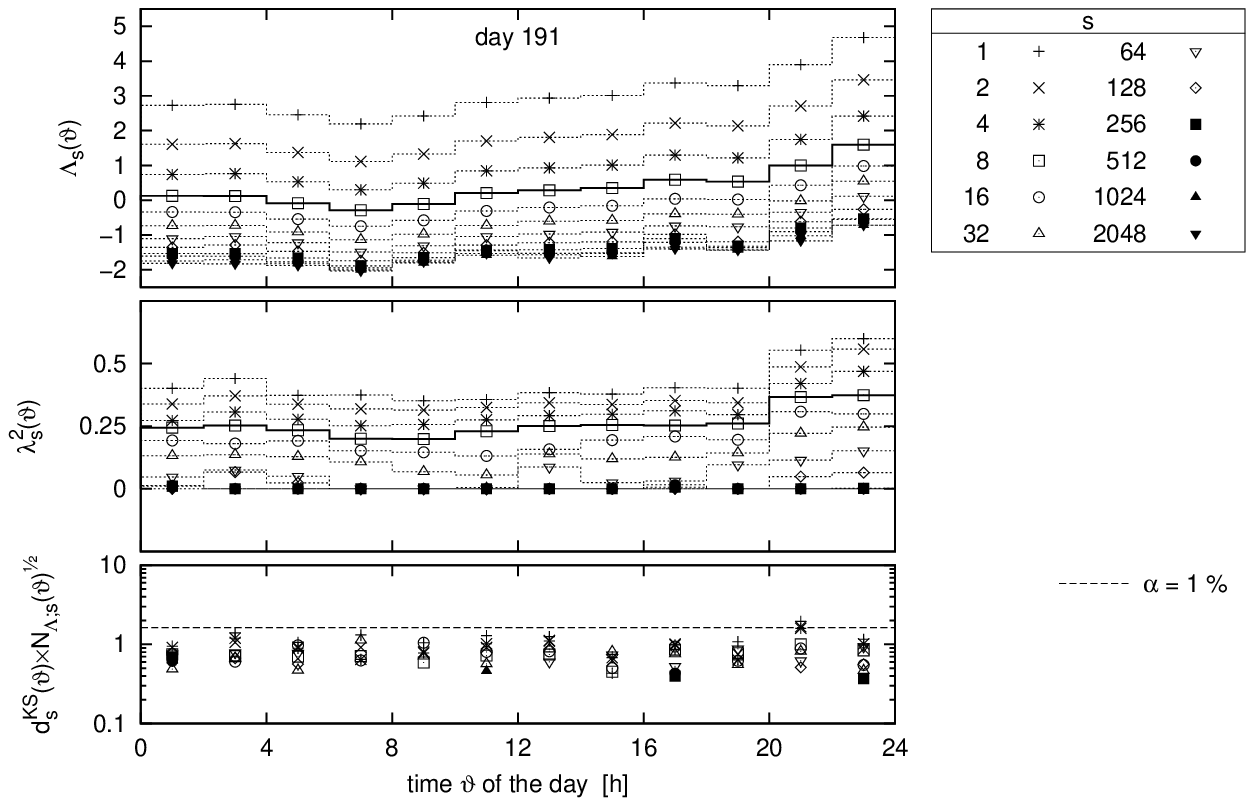}
	}
	\caption{
	  The top (central) panel shows the mean (variance) of 
	  $\Lambda$ for each sub-sample of the data acquired at day 191 at the
	  \cite{lammefjord} site and for several $s$.
	  The mean and variance can be identified with the position and 
	  shape parameter, respectively.
	  The solid line corresponds to $s=8$ and is therefore related to the
	  graphs in figure~\ref{F:super:day191:24h}. 
	  The test value of the Kolmogorov-Smirnov test, 
	  $d^{\rm KS}_s(\vartheta)\times\sqrt{N_{\Lambda;s}(\vartheta)}$,
	  is plotted in the lowest panel. The dashed line corresponds
	  to the critical value for the significance level $\alpha=1~\%$
	  }
	\label{F:super:day191:2h}
\end{figure}
 
\cite{boettcher07} showed that the distribution of ABL wind speed
increments can be understood as a superposition of different subsets of
isotropic turbulence.  Indeed, the depicted $\Lambda$-series in the lowest
panel of figure~\ref{F:super:day191:24h} indicates that the hypothesis
``$\Lambda$ is of a normal distribution'' might be fulfilled on a smaller
period than 24~h.  Hence, the 24~h increment series $(x_{s;n})_{n=0}^{N-1}$ was
divided into twelve 2~h sub-samples where each of which represents a time
$\vartheta$ of the day. Each sub-sample was analysed with respect to
superstatistics individually.  After computing and comparing the time scales
$T_s(\vartheta)$ and $\tau_s(\vartheta)$, the $\Lambda$-series for each
sub-sample around $\vartheta$ was extracted and tested for the hypothesis of
being normally distributed. This test was done using a Kolmogorov-Smirnov test
\cite[see e.g.][chapter 8]{daniel90}.  Its test value corresponds to the
largest deviation of the empirical cumulative distribution function (c.d.f.)
$S_{s}(\Lambda,\vartheta) $ from the corresponding Gaussian c.d.f.\
$\Phi_0\left(\frac{\Lambda-\Lambda_s(\vartheta)}{\lambda_s(\vartheta)}\right)$
with fitted $\Lambda_s(\vartheta)$ and $\lambda_s^2(\vartheta)$. It is denoted
by
\begin{equation} \label{E:super:ks}
 	d^{\rm KS}_s(\vartheta) 
	= \sup_{\Lambda} 
	\left| 
		S_{s}(\Lambda,\vartheta) -
	  	\Phi_0\left(
			\frac{\Lambda-\Lambda_s(\vartheta)}{\lambda_s(\vartheta)}
		\right) 
	\right|.
\end{equation} 
The hypothesis $H_0$ in~\eqref{E:super:hl} is rejected on a significance
level $\alpha$ if 
\begin{equation} \label{E:super:ks2}
 	d^{\rm KS}_s(\vartheta) \times \sqrt{N_{\Lambda;s}(\vartheta)}
	\ge 
	d^{\rm KS}_{\rm crit}(\alpha).
\end{equation} 
The critical value is 1.63 for $\alpha=1\,\%$ and $N_{\Lambda;s}(\vartheta)>40$
where the latter denotes the number of data points in the $\Lambda$-series for
increment length $s$ and time $\vartheta$ \cite[see e.g.][table~A.18 for
tabulated values]{daniel90}.

Figure~\ref{F:super:day191:2h} shows the  results for each 2~h sub-sample of the
Lammefjord day 191 data plotted as a function of time $\vartheta$ for a variety
of increment lengths $s$. The top and central panel depict the mean and variance
of $\Lambda$ for each sub-sample, respectively. According
to~\eqref{E:super:fit1} and \eqref{E:super:fit2} they can be identified
with $\Lambda_s=\ln\beta_s$ and
$\lambda_s^2$, respectively. The bottom panel shows the test value of the
Kolmogorov-Smirnov test and the critical value for the significance level
$\alpha=1~\%$. The graph allows the conclusion that the hypothesis ``the
$\Lambda$-series of each sub-sample around $\vartheta$ and 
for different $s$ is of a normal
distribution'' can not be rejected on a significance level $\alpha=1~\%$.
Moreover, the
$\lambda_s^2$-plots in the central panel reveal that for large $s$ the shape
parameter of each sub-sample approaches zero as it is expected from the
intermittency hypothesis. Therefore, it can be concluded that 
Castaing's hypothesis is fulfilled during time intervals of 2~h. But it is
violated on larger time scales, such as 24~h, due to non-stationarity which 
are reflected by the time dependence of $\Lambda_s(\vartheta)$ and
$\lambda_s^2(\vartheta)$.  In other words, the distribution
shown in figure~\ref{F:super:day191:24h} is a superposition of Gaussians with
different means $\Lambda_s(\vartheta)$ and variances $\lambda_s^2(\vartheta)$ for
$s=8$ and is thus not exactly a normal shaped distribution.

\begin{figure}
	\centerline{
	  \includegraphics{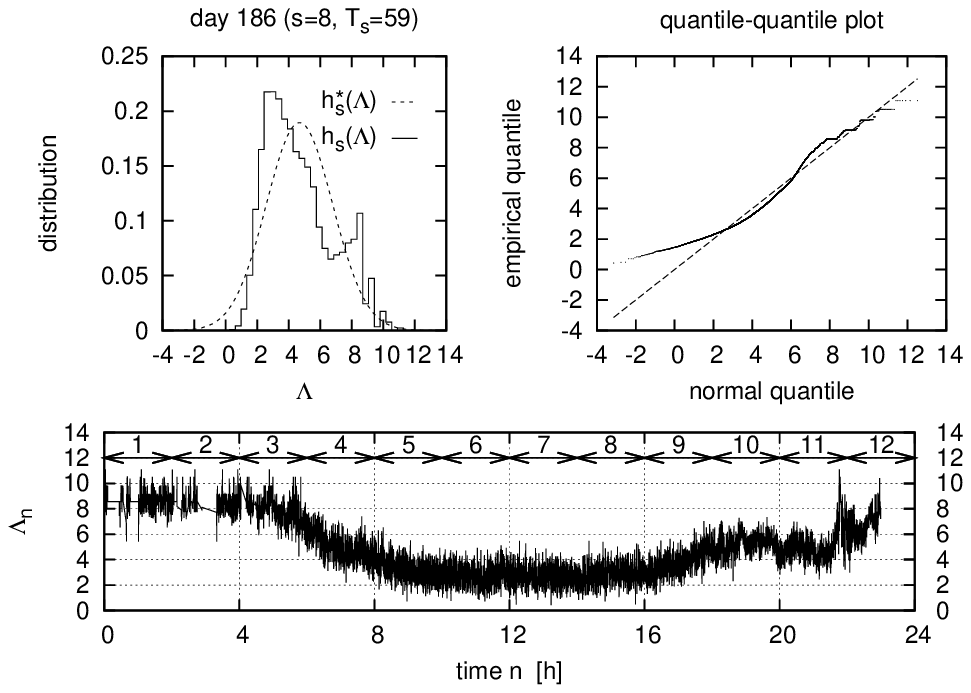}
	}
	\caption{
	  The top left panel displays the estimated 24~h $\Lambda$-distribution
	  $h_s(\Lambda)$ for $s=8$ of the \cite{lammefjord} measurement day 186.
	  $h_s^*(\Lambda)$ denotes a Gaussian distribution with mean
	  $\mean{\Lambda}$ and variance $\mean{(\Lambda-\mean{\Lambda})^2}$.  The
	  quantile-quantile plot is shown in the upper right panel. The bottom
	  panel displays the $\Lambda$-series for that day. The numbers 1 to 12
	  illustrate the twelve 2~h sub-samples which are individually investigated
	  with respect to superstatistics
	}
	\label{F:super:day186:24h}
\end{figure}
\begin{figure}
	\centerline{
	  \includegraphics{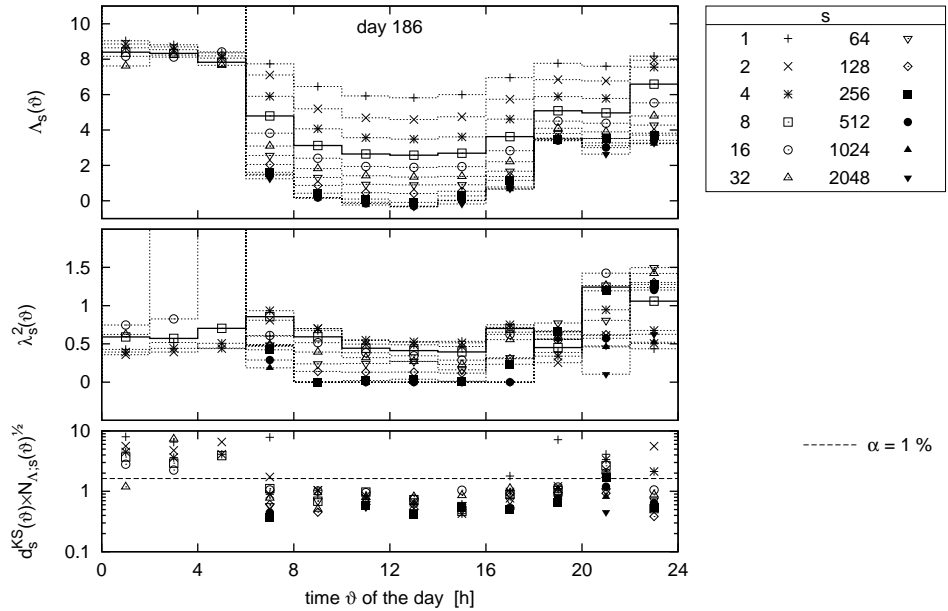}
	}
	\caption{
	  The top (central) panel shows the mean (variance) of 
	  $\Lambda$ for each sub-sample of the data acquired at day 186 at the
	  \cite{lammefjord} site and for several $s$.
	  The mean and variance can be identified with the position and 
	  shape parameter, respectively.
	  The solid line corresponds to $s=8$ and is therefore related to the
	  graphs in figure~\ref{F:super:day186:24h}. 
	  The test value of the Kolmogorov-Smirnov test, 
	  $d^{\rm KS}_s(\vartheta)\times\sqrt{N_{\Lambda;s}(\vartheta)}$,
	  is plotted in the lowest panel. The dashed line corresponds
	  to the critical value for the significance level $\alpha=1~\%$
	}
	\label{F:super:day186:2h}
\end{figure}
The same analysis was done with the Lammefjord day 186 data which did not show
a clear cross-over behaviour in figure~\ref{F:inc:hi}. The superstatistical
algorithm was used to extract the $\Lambda$-series from the 24~h time series
$(x_{s;n})_{n=0}^{N-1}$
and tested for Gaussianity. The upper left panel of
figure~\ref{F:super:day186:24h} shows the histogram $h_s(\Lambda)$ for
$s=8$ which is clearly non-Gaussian shaped. 
This  explains the small deviation of $p_s(x_s)$ in
figure~\ref{F:inc:hi} from the fitted Castaing distribution for $s=8$.
The 24~h increment series $(x_{s;n})_{n=0}^{N-1}$ was also divided into twelve
2~h sub-samples where each of which was analysed with respect
to superstatistics individually. Figure~\ref{F:super:day186:2h} 
shows the Castaing parameters $\Lambda_s(\vartheta)$ and
$\lambda^2_s(\vartheta)$ as a function of time $\vartheta$. The time span
$8~{\rm h}\lesssim \vartheta \lesssim 18~{\rm h}$ is the only
region where Castaing's hypothesis can not be rejected
on a 2~h scale with $\alpha=1~\%$: the Kolmogorov-Smirnov test value
$d^{\rm KS}_s(\vartheta)\times\sqrt{N_{\Lambda;s}(\vartheta)}$ is below
the critical value and the shape parameter $\lambda^2_s(\vartheta)$ goes to
zero as $s$ gets larger. That means that both Castaing's hypothesis and the
cross-over behaviour can be recovered in the mentioned time span  on windows of
2~h period. Outside this region, the length of 2~h for the sub-samples is still
too large or in other words the resolution is too low for recovering a normally
distributed $\Lambda$-series. However, going to smaller sub-samples gives worse
statistics due to the reduced number of data points in each sub-sample.
Nevertheless, the increments of this day are much more non-stationary than the
data gathered at day 191. It exemplifies that high non-stationarity of wind speed
data cause a fat tailed increment distribution for large increment length $s$.

The data acquired at day 192 at the Lammefjord site 
show similar behaviour to the day 191 data when analysed
with respect to superstatistics.

 
As a summary, the superstatistical approach is sensitive enough to
detect sub-regions of the increment series where  Castaing's
hypothesis is fulfilled. In such a region the increment
distribution takes the shape given in~\eqref{E:castaing2}.
The position and shape parameter change with time as they  differ between
different sub-regions. This makes this
approach be capable of determining the dynamics of
$\Lambda_s(\vartheta)=\ln\beta_s(\vartheta)$ and $\lambda^2_s(\vartheta)$.
Moreover, it can be used to verify that wind speed data with large fluctuating
Castaing parameters have a non-Gaussian increment distribution at large increment lengths.


\section{Conclusions}
 
Our study verified the picture of natural ABL turbulence to be
composed of successively occurring close to ideal turbulence with different
parameters.
  
We showed that in good approximation the fluctuation of the the wind speed
around its window average $V$ is of a symmetric normal distribution with a
standard deviation being proportional to $V$.  The proportionality factor can
be estimated by the standard deviation of the normalised fluctuation. The
investigation of the time dependent volatility of the normalised fluctuation
leads to the time resolution of the proportionality factor. Our analysis showed
that approximating it by being constant over 24~h is reasonable. However,
within 24~h the mean wind speed changed frequently leading to two separate time
scales: the time scale on which the mean wind speed changes and the time scale
on which the proportionality factor between the mean wind speed and the
standard deviation of the fluctuation can be approximated as being constant.
The mean speed of stationary laboratory turbulence does not alter with
time so that the $V$ dependence of the fluctuation distribution can not be
investigated by means of such experiments.
Nevertheless, the fluctuation is of a normal distribution, too, leading to the
conclusion that ABL turbulence is a sequence of stationary turbulence with
varying mean. 
 
The intermittency behaviour of ABL wind speed was tested using the increment
statistics approach.  The coincidence between the empirical histograms and the
hypothetical distributions makes the increment series in this representation
look stationary. However, Castaing's intermittency hypothesis involves a time
scale separation. On a small scale the wind speed increments are of a symmetric
normal distribution whose variance alters on a larger scale according to a
log-normal distribution.

We checked this time scale separation using a superstatistical approach. We
found that there is a ``critical'' time scale below which the increment series
behaves normally distributed and above which the increment series is of a
leptokurtic distribution.  In the terminology of superstatistics this time
scale is referred to as the large time scale in contrast to the small time
scale reflecting the relaxation time of the increment process.  If the latter
is small compared to the large time scale it is possible to estimate the
variance of each Gaussian segment and analyse their statistics. We found that
their logarithm is not exactly of a stationary normal distribution but rather
of a sequence of normal distributions with varying mean and variance. This
incorporates a third time scale into the ABL wind speed increment series on
which the log-normal distribution of the variances in the sense of Castaing's
hypothesis can be approximated as being stationary.

We additionally verified that highly non-stationary turbulence 
might show a fat tailed increment distribution even at large
increment length $s$.


\paragraph{Acknowledgements.}
The study was supported by Germany's Federal Ministry for Education and
Research (BMBF) under grant number 03SF0314. It is part of the joint project
``Statistical analysis and stochastic modelling of turbulent gusts in surface
wind''.
 


\begin{thebibliography}{}

\bibitem[Beck et~al., 2005]{beck05_b}
Beck, C., Cohen, E., and Swinney, H. (2005).
\newblock From time series to superstatistics.
\newblock {Phys. Rev. E}, 72:056133.

\bibitem[Beck and Cohen, 2003]{BC03}
Beck, C. and Cohen, E. G.~D. (2003).
\newblock Superstatistics.
\newblock {Physica A}, 322:267--275.

\bibitem[Boettcher et~al., 2007]{boettcher07}
Boettcher, F., Barth, S., and Peinke, J. (2007).
\newblock Small and large scale fluctuations in atmospheric wind speeds.
\newblock {Stoch. Environ. Res. Risk Assess.}, 21:299--308.

\bibitem[Boettcher et~al., 2003]{boettcher02}
Boettcher, F., Renner, C., Waldl, H.~P., and Peinke, J. (2003).
\newblock On the statistics of wind gusts.
\newblock {Bound.-Layer Meteor.}, 108:163--173.

\bibitem[Burton et~al., 2004]{we_handbook}
Burton, T., Sharpe, D., Jenkins, N., and Bossanyi, E. (2004).
\newblock {Wind Energy Handbook}.
\newblock John Wiley.

\bibitem[Castaing et~al., 1990]{CGH90}
Castaing, B., Gagne, Y., and Hopfinger, E.~J. (1990).
\newblock Velocity probability density-functions of high {Reynolds} number
  turbulence.
\newblock {Physica D}, 46:177--200.

\bibitem[Daniel, 1990]{daniel90}
Daniel, W. (1990).
\newblock {Applied Nonparametric Statistics}.
\newblock PWS-Kent, second edition.

\bibitem[Kolmogorov, 1941a]{kolmogorov41c}
Kolmogorov, A.~N. (1941a).
\newblock Dissipation of energy in locally isotropic turbulence.
\newblock {Dokl. Akad. Nauk SSSR}, 32:16--18.

\bibitem[Kolmogorov, 1941b]{kolmogorov41a}
Kolmogorov, A.~N. (1941b).
\newblock The local structure of turbulence in incompressible viscous fluid for
  very large {Reynolds} number.
\newblock {Dokl. Akad. Nauk SSSR}, 30:299--303.

\bibitem[Kolmogorov, 1962]{kolmogorov62}
Kolmogorov, A.~N. (1962).
\newblock A refinement of previous hypotheses concerning the local structure of
  turbulence in a viscous incompressible fluid at high {Reynolds} number.
\newblock {J. Fluid Mech.}, 13:82--85.

\bibitem[Lammefjord, 1987]{lammefjord}
Lammefjord (1987).
\newblock Lammefjord data obtained from the {Ris\o\ National Laboratory} in
  {Denmark}, {\tt http://www.risoe.dk/vea}, through {\tt
  http://www.winddata.com}.

\bibitem[Obukhov, 1962]{obukhov62}
Obukhov, A. (1962).
\newblock Some specific features of atmospheric turbulence.
\newblock {J. Fluid Mech.}, 13:77--81.

\bibitem[Queiros, 2007]{queiros07}
Queiros, S. M.~D. (2007).
\newblock On new conditions for evaluate long-time scales in superstatistical
  time series.
\newblock {Physica A}, 385:191--198.

\bibitem[Renner et~al., 2001]{freejet}
Renner, C., Peinke, J., and Friedrich, R. (2001).
\newblock Experimental indications for {Markov} properties of small-scale
  turbulence.
\newblock {J. Fluid Mech.}, 433:383--409.

\bibitem[Taylor, 1938]{taylor38}
Taylor, G. (1938).
\newblock The spectrum of turbulence.
\newblock {Proc. R. Soc. Lond. {\rm A}}, 164:476--490.

\bibitem[Wallace and Hobbs, 2006]{wallace06}
Wallace, J.~M. and Hobbs, P.~V. (2006).
\newblock {Atmospheric Science}.
\newblock Academic Press Elsevier, second edition.

\end{thebibliography}
\end{document}